\documentclass[aps,prl,twocolumn,showpacs,superscriptaddress,footinbib]{revtex4-1} 
\usepackage{amssymb}
\usepackage{amsmath}
\usepackage[colorlinks=true,linkcolor=blue,citecolor=blue, urlcolor=blue]{hyperref} 
\usepackage{tikz}
\usepackage{graphicx}
\usepackage[utf8]{inputenc}
\definecolor{uibred}{RGB}{167, 38, 47}

\newcommand{\vp}{\mathbf{p}}
\newcommand{\vpt}{|\mathbf{p}|}
\newcommand{\mysection}[1]{{\vspace{10 pt}\noindent \emph{{#1}.--}}}

\newcommand{\rff}[1]{Fig.~\ref{#1}}

\begin{document}
	
	\title{Stable and unstable perturbations in universal scaling phenomena far from equilibrium} 
	
	\author{Thimo Preis} \email{preis@thphys.uni-heidelberg.de}
	\affiliation{Institut f\"{u}r Theoretische Physik, Universit\"{a}t Heidelberg, 69120 Heidelberg, Germany}
	
	\author{Michal P.\ Heller} \email{michal.p.heller@ugent.be}
	\affiliation{Department of Physics and Astronomy, Ghent University, 9000 Ghent, Belgium}
	
	\author{J{\"u}rgen Berges} \email{berges@thphys.uni-heidelberg.de}
	\affiliation{Institut f\"{u}r Theoretische Physik, Universit\"{a}t Heidelberg, 69120 Heidelberg, Germany}

	\begin{abstract}
		We study the dynamics of perturbations around nonthermal fixed points associated to universal scaling phenomena in quantum many-body systems far from equilibrium. For an $N$-component scalar quantum field theory in 3+1 space-time dimensions, we determine the stability scaling exponents using a self-consistent large-$N$ expansion to next-to-leading order.
		Our analysis reveals the presence of both stable and unstable perturbations, the latter leading to quasi-exponential deviations from the fixed point in the infrared. We identify a tower of far-from-equilibrium quasi-particle states and their dispersion relations by computing the spectral function. With the help of linear response theory, we demonstrate that unstable dynamics arises from a competition between elastic scattering processes among the quasi-particle states. What ultimately renders the fixed point
		dynamically attractive is the phenomenon of a \textit{scaling
			instability}, which is the universal scaling of the unstable regime
		towards the infrared due to a self-similar quasi-particle cascade. 
		Our results provide ab initio understanding of emergent stability properties in self-organized scaling phenomena.
	\end{abstract}
	
	\maketitle
	
	\mysection{Introduction} 
	One of the greatest challenges in physics is to understand how emergent collective behavior arises from the underlying quantum dynamics of a system. A~prominent example concerns universal scaling phenomena associated to phase transitions after fine-tuning of relevant parameters, such as adjustment to a critical temperature~\cite{Hohenberg:1977ym}. By contrast, scaling phenomena without fine-tuning of parameters can play an important role for the build-up of complex structures, in particular, far from equilibrium~\cite{Bak:1987xua,bak_how_1996}.
	On a fundamental quantum level important examples include universal scaling in quantum many-body systems associated to nonthermal fixed points, which exhibit attractor properties~\cite{Berges:2020fwq}. Here the applications range from relativistic collisions with heavy nuclei~\cite{Berges:2013eia,Kurkela:2015qoa}, and early universe cosmology~\cite{Micha:2002ey,Berges:2008wm}, to experiments with ultracold quantum gases providing unprecedented data for universal scaling in isolated~\cite{Prufer:2018hto,Erne:2018gmz,Glidden:2020qmu}, as well as driven systems~\cite{Navon:2016em,Helmrich:2020sgn}.
	However, an understanding of the observation of attractor behavior for scaling solutions based directly on the underlying quantum dynamics is still missing.   
	
	In this work, we compute the stability properties of a nonthermal fixed point for an interacting scalar quantum field theory in 3+1 space-time dimensions from first principles. While the approach to universal scaling behavior of this system is known to be observed from a wide range of far-from-equilibrium initial conditions without fine-tuning~\cite{Berges:2014bba,PineiroOrioli:2015cpb,Chantesana:2018qsb,Deng:2018xsk,PineiroOrioli:2018hst,Boguslavski:2019ecc}, our analysis reveals the presence of both stable and unstable perturbations around the scaling solution. The unstable modes lead to quasi-exponential deviations from the fixed point in the infrared. This is exemplified in Fig.~\ref{fig:deltaF_intro}, where the time evolution of the statistical correlation function $F$ (see below) for two different, unstable (red) and stable (blue), momenta is shown.
	
	Our results are obtained from a systematic large-$N$ expansion to next-to-leading order, where $N$ is the number of components of the real field operator $\hat{\Phi}_{a=1,\ldots N}(t,\mathbf{x})$ as a function of time $t$ and space~$\mathbf{x}$ or, respectively, spatial Fourier momentum $\mathbf{p}$. The full unitary quantum dynamics of correlation functions at this order is obtained numerically. We uncover the underlying physical processes using a linear response analysis around the time-evolving scaling solution, and explain the observations of attractive behavior at fixed momenta in the presence of both negative and positive stability exponents.
	\begin{figure}[t!]
		\centering
		\includegraphics[width=1\columnwidth]{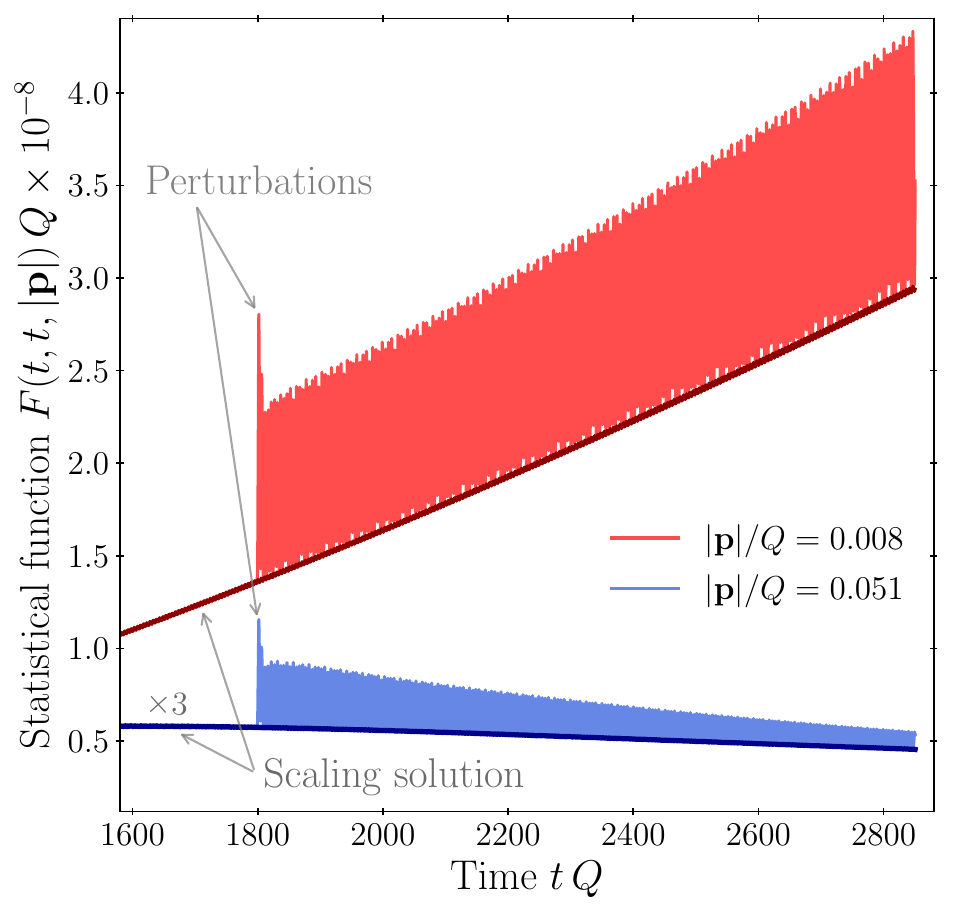}
		\caption{Statistical function $F(t,t,\vpt)$ for two different values of the momentum $\vpt$ as a function of time. Shown are the unperturbed scaling solutions (solid bold lines), and the responses after a perturbation $\delta F(t_i,t_i,\vpt)$ at time $t_i\,Q=1800$. We find that perturbations grow (red) at sufficiently low momenta and decay (blue) at higher momenta.
		}
		\label{fig:deltaF_intro}
	\end{figure}
	
\begin{figure}[t!]
		\centering
		\includegraphics[width=1\columnwidth]{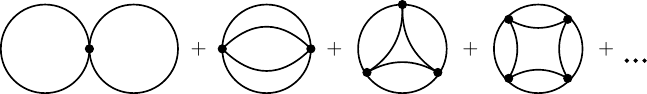}
		\caption{Diagrammatic representation of the infinite series of contributions to $\langle\hat{H}(t)\rangle$ taken into account with the $2$PI effective action to NLO in the large-$N$ expansion. Lines correspond to self-consistently resummed propagators and circles to bare vertices.	}
		\label{fig:diagrams}
	\end{figure}

	\mysection{Far-from-equilibrium quantum fields}
	We consider a relativistic quantum field theory described by the $O(N)$-symmetric Hamilton operator
	\begin{align}
	\hat{H}(t) = \int d^3x &\left[\frac{1}{2} \hat{\Pi}^2_a(t,\mathbf{x}) + \frac{1}{2}\left( \nabla_{\mathbf{x}} \hat{\Phi}_a(t,\mathbf{x})\right)^2 \right.\nonumber\\
	&\left.+  \frac{\lambda}{4!N} \left(\hat{\Phi}_a(t,\mathbf{x}) \hat{\Phi}_a(t,\mathbf{x})\right)^2 \right]
	\end{align}
	with $\hat{\Pi}_a(t,\mathbf{x}) = \partial_t \hat{\Phi}_a(t,\mathbf{x})$ and a summation over repeated indices is implied. While we specify values of the coupling $\lambda$ for the massless fields below, the detailed values of these parameters do not enter universal quantities~\cite{PineiroOrioli:2015cpb}. We solve for the nonequilibrium dynamics employing a self-consistent large-$N$ expansion to next-to-leading order~\cite{Berges:2001fi,Aarts:2002dj}. Diagrammatically, this takes into account an infinite series of quantum corrections, which is indicated in Fig.~\ref{fig:diagrams} for the expectation value of the Hamiltonian. More precisely, we solve the quantum evolution equations for the connected correlation functions, i.e.\ the expectation value of the anti-commutator (statistical function) 
	\begin{eqnarray}
	F_{ab}(t,t',{\mathbf x}-{\mathbf x'}) &=& \frac{1}{2}\langle \{\hat{\Phi}_a(t,{\mathbf x}),\hat{\Phi}_b(t',{\mathbf x'})\}\rangle
	\nonumber\\
	&& -\langle \hat{\Phi}_a(t,{\mathbf x})\rangle \langle\hat{\Phi}_b(t',{\mathbf x'})\rangle
	\end{eqnarray}
	and of the commutator (spectral function)
	\begin{equation}
	\label{eq.rhodef}
	\rho_{ab}(t,t',{\mathbf x}-{\mathbf x'}) = i\langle [\hat{\Phi}_a(t,{\mathbf x}),\hat{\Phi}_b(t',{\mathbf x'})]\rangle 
	\end{equation}
	for spatially homogeneous and isotropic systems.
	The nonequilibrium time evolution equations are derived from the two-particle irreducible (2PI) effective action on the closed time contour at next-to-leading order in the resummed large-$N$ expansion.
	By virtue of $O(N)$ symmetry we consider $\langle \hat{\Phi}_a(t,{\mathbf x}) \rangle =0$, such that the correlators are diagonal in field space with $F_{ab}=\delta_{ab}\, F$ and $\rho_{ab}=\delta_{ab}\, \rho$.
	
	The spectral function is related to the retarded propagator as $G^R(t,t',{\mathbf x}) = \theta(t-t') \rho(t,t',{\mathbf x})$ and encodes the bosonic equal-time commutation relations $\rho(t,t',{\mathbf x}) |_{t=t'} = 0$, $\partial_{t} \rho(t,t',{\mathbf x}) |_{t=t'} = \delta^{(3)}(\mathbf{x})$. While the latter determine also $\rho$ at some initial time $t_0$, for $F$ we consider the following class of initial conditions in spatial Fourier space: 
	\begin{eqnarray}
	\label{eq:initialcond}
	F(t_0,t_0,\vp) = \frac{1}{\sqrt{\vp^2+m^2}}\left[\left(\frac{N n_0}{\lambda}\right) \theta(Q-\vpt) +\frac{1}{2}\right]\,\,\,\;
	\end{eqnarray}
	in the limit $m^2 \to 0^+$.
	To start the evolution far from equilibrium, we choose $N n_0/\lambda\gg 1$ up to the momentum scale $Q$ determining the (conserved) energy density of the system.
	
	\mysection{Perturbations around the universal scaling solution}
	The subsequent time evolution approaches self-similar scaling behavior described by
	\begin{equation}
	\label{eq:scaling_F_equaltime}
	F(t,t,|\vp|) = (t/t_{\mathrm{ref}})^{\alpha} F_S\left((t/t_{\mathrm{ref}})^\beta |\vp|\right)
	\end{equation}
	corresponding to the solid lines exemplified in Fig.~\ref{fig:deltaF_intro}. The scaling exponents are universal, as well as the form of the scaling function $F_S$ depending on the single scaling variable $(t/t_{\mathrm{ref}})^\beta |\vp|$, with $\beta \simeq 1/2$, $\alpha = d\beta$ and $d=3$ in our case~\cite{PineiroOrioli:2015cpb}. Here $t_{\mathrm{ref}}$ denotes some reference time, which does not affect universal properties and is described below.  
	
	The approach to (\ref{eq:scaling_F_equaltime}) has been studied starting from a wide range of different initial conditions. Examples include parametric resonance initial conditions in a strong field regime
	\cite{Berges:2008wm,Berges:2016nru} or large initial occupation numbers similar to (\ref{eq:initialcond}) \cite{Berges:2014bba,PineiroOrioli:2015cpb,Mikheev:2018adp,Deng:2018xsk,PineiroOrioli:2018hst,Boguslavski:2019ecc}. Corrections to the scaling exponents of the high-momentum cascade were recently explored via kinetic methods~\cite{Mikheev:2022fdl,Brewer:2022vkq} in the context of the prescaling phenomenon~\cite{Mazeliauskas:2018yef}.
	Furthermore, nonthermal fixed points were realized in cold atom experiments, e.g., from sudden quenches across phase transitions~\cite{Prufer:2018hto,Erne:2018gmz,Glidden:2020qmu}. While universal quantities like scaling exponents do not depend on microscopic details, such as specific values of $\lambda$ or $n_0$, the time for the approach to scaling is not a universal quantity. Moreover, scaling according to (\ref{eq:scaling_F_equaltime}) is a transient phenomenon in an isolated system, with the system eventually approaching thermal equilibrium~\cite{Berges:2016nru}. The latter can always be delayed by increasing $(N n_0)/\lambda$ entering the initial condition (\ref{eq:initialcond}) such that (\ref{eq:scaling_F_equaltime}) is accurately realized for suitable time ranges in units of~$Q$. 
	
	For the numerical results presented in this work, we employ $\lambda=0.01$, $n_0=25$ and $N=4$, which e.g.\ reflects the Higgs sector of the Standard Model of particle physics~\footnote{Temporal and spatial grid points are respectively discretised with the spacing $a_t\,Q=0.3$ and $a_x\,Q=0.75$ with $N_x=495$ spatial grid points.
		For the numerical parameters explored in this work, we have checked that the relevant physical results are insensitive to changes in the IR and UV cutoffs, as well as the temporal spacing. We have also explicitly varied $N$ to other values, such as $N=8$, to check for the robustness of our results. The code is available at \cite{shen:2020slv}.}.
	Starting from the initial conditions we consider, the universal scaling regime of interest emerges in the infrared after times of order $10^3/Q$ and our scaling solution results are in accordance with earlier studies of nonthermal fixed points. 
	
	After the behavior~\eqref{eq:scaling_F_equaltime} is established, we consider perturbations around the scaling solution according to
	\begin{equation}
	\label{eq.Fpert}
	F(t,t,\vpt) = (t/t_{\mathrm{ref}})^{\alpha} F_S\left((t/t_{\mathrm{ref}})^\beta \vpt\right) + \delta F(t,t,\vpt)
	\end{equation}
	by suddenly turning on $\delta F > 0$ at a given time $t_i$. Perturbations of two different momentum modes at $t_i\,Q=1800$ are depicted in Fig.~\ref{fig:deltaF_intro}. In the following we analyze the underlying processes that lead to the quasi-exponential growth of perturbations in time at low values of $\vpt$ and decay at higher momenta. 
	
	\begin{figure*}[t!]
		\centering
		\includegraphics[width=1\linewidth]{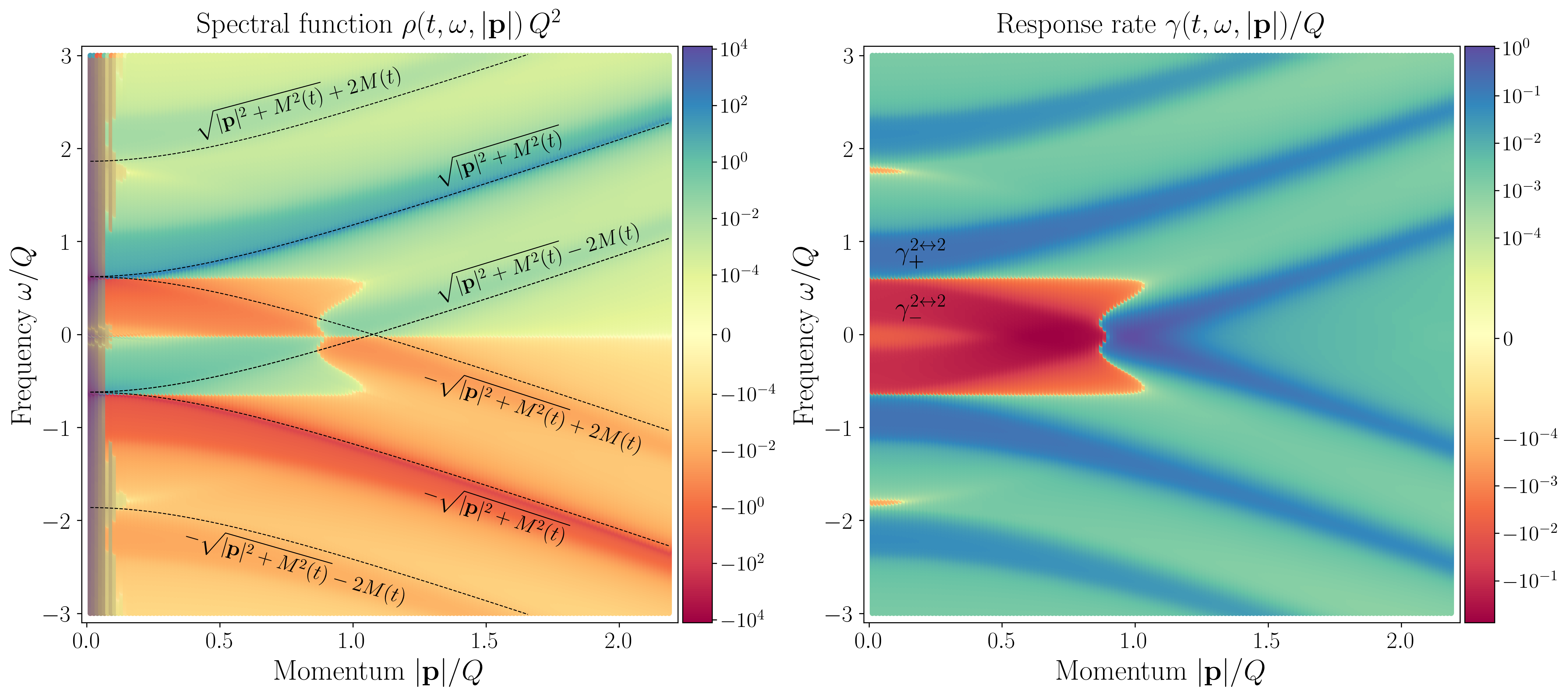}
		\caption{Left: Frequency and momentum dependence of the nonequilibrium spectral function $\rho(t,\omega,|\vp|) = -\rho(t,-\omega,|\vp|)$ at $t\,Q = 2400$. Apart from the quasi-particle peak with relativistic dispersion relation, we find a tower of additional quasi-particle peaks. Right: Linear response rate $\gamma(t,\omega,|\vp|) = \gamma(t,-\omega,|\vp|)$ versus frequency and momentum at the same time. For low momenta, the on-shell response rate at $\omega \simeq M$ receives competing contributions from adjacent peaks and turns negative.
		}
		\label{fig:decay_wig}
	\end{figure*}
	
	\mysection{Far-from-equilibrium quasi-particles} Since the universal exponent $\beta$ is positive, the scaling behavior (\ref{eq:scaling_F_equaltime}) describes an inverse cascade towards low momenta during which a macroscopically occupied zero-momentum mode emerges~\cite{PineiroOrioli:2015cpb}. We demonstrate in the following that this gives rise to a tower of emergent quasi-particles excitations out of equilibrium. The spectrum is encoded
	in the spectral function $\rho$, which we access by a Wigner transform with respect to relative time $\Delta t=t-t^\prime$. 
	Expressed as a function of the central time $\tau=\tfrac{1}{2}(t+t^\prime)$, the spectral function is obtained via (a factor of $i$ is included for the Wigner transformed $\rho$ to ensure realness)
	\begin{equation}
	\label{eq.rhoomegap}
	\rho(\tau,\omega,\vpt) =-i \int_{-2\tau}^{2\tau} d\Delta t\, e^{i \omega \Delta t} \rho(\tau+\tfrac{\Delta t}{2}, \tau-\tfrac{\Delta t}{2},\vpt).
	\end{equation}
	For notational simplicity, we write $\tau=t$ in the following, and the Wigner transformed spectral function is shown in the left graph of~\rff{fig:decay_wig} for $t\, Q = 2400$. 
	One observes a pronounced quasi-particle peak obeying a relativistic dispersion relation $\omega(t,\vp) \equiv \sqrt{\vp^2+M^2(t)}$ with in-medium mass $M(t)$. Its value is established quickly and approximately constant in the scaling regime with $M/Q \simeq 0.62$ for the parameters employed for~\rff{fig:decay_wig}. The associated widths scale to zero in the scaling regime as time progresses, leading to increasingly long-lived quasi-particles~\cite{PineiroOrioli:2018hst,Boguslavski:2019ecc}.
	
	However, we also find additional quasi-particles to be present. Their dispersion relations are well described by
	\begin{equation}
	\omega^\pm(t,\vp)=2M(t) \pm  \omega(t,\vp) \, .
	\end{equation}
	This multitude of excitations out of equilibrium may be understood as phase fluctuations of a macroscopic zero mode rotating in $N$-component field space~\cite{Boguslavski:2019ecc}, here built up by the inverse cascade.
	Though the additional quasi-particle modes at $\omega^\pm(t,\vp)$ exhibit a much lower magnitude as compared to the pronounced peak at $\omega(t,\vp)$, their contributions play a leading role for the stability properties of the system at sufficiently low momenta, as we show next. 
	
	\begin{figure}[t!]
		\centering
		\includegraphics[width=1\columnwidth]{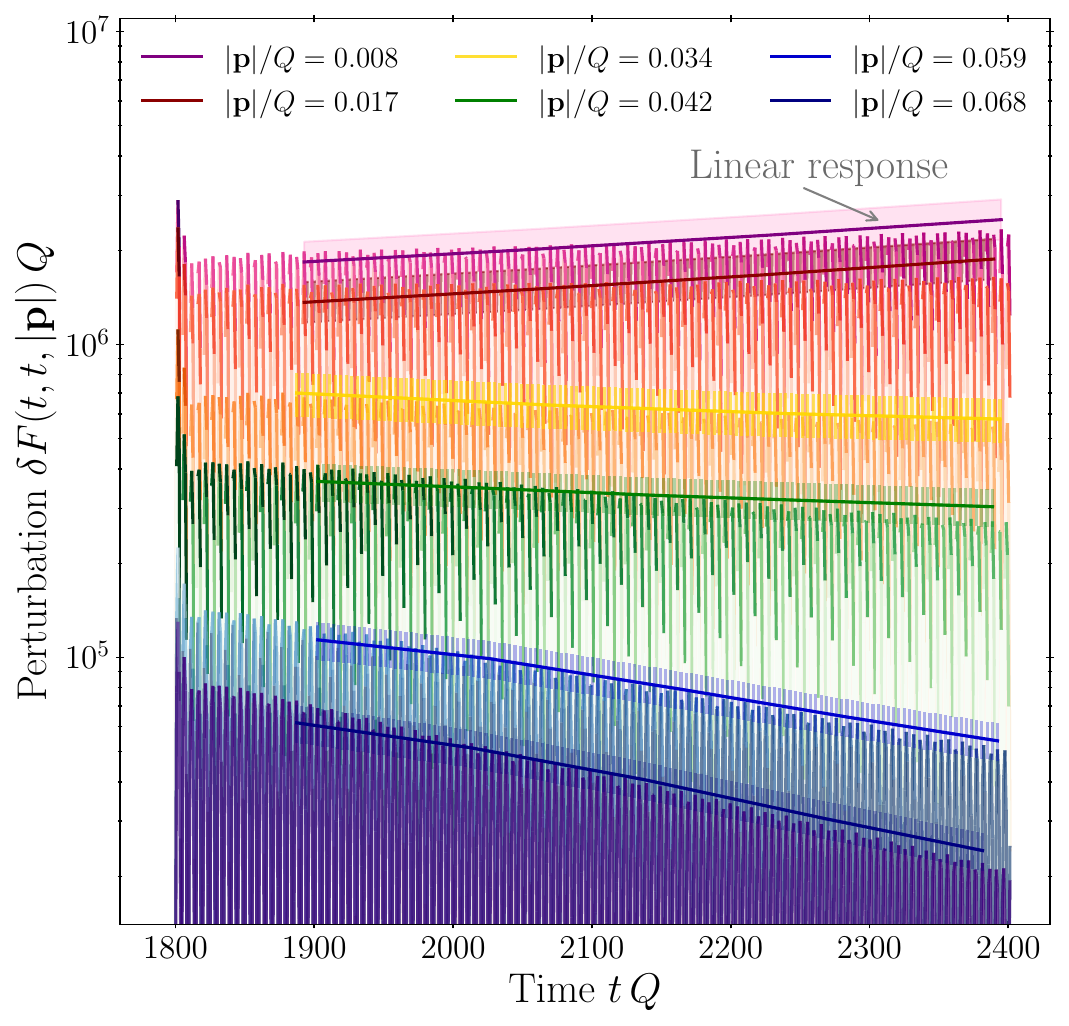}
		\caption{The perturbation $\delta F(t,t,\vpt)$ around the universal scaling solution as a function of time for different momenta. The growth and decay of the envelopes for the oscillatory full solution of the quantum evolution equations is rather well described by the linear response result.}
		\label{fig:deltaF_comparison}
	\end{figure}
	
	\mysection{Linear response around the universal scaling solution} To gain further analytical insights, we perform a linear response analysis around the nonequilibrium scaling solution, described by 
	\begin{equation}
	\label{eq:lr_diagonal}
	\partial_t	\delta F(t,\omega,\vpt) = -\gamma(t,\omega,\vpt)\, \delta F(t,\omega,\vpt) \, .
	\end{equation}
	Here $\gamma(t,\omega,\vpt)$ denotes the {\it time-dependent} response rate for the frequency- and momentum-resolved (Wigner transformed) perturbations of the statistical function. This response rate for the far-from-equilibrium state is time dependent, in contrast to conventional linear response estimates around equilibrium solutions. The dynamics of the perturbation
	\begin{equation}
	\label{eq:lr_diagonal_solution}
	\delta F(t,\omega,\vpt) = \delta F(t_i,\omega,\vpt) \exp\left[-\Gamma(t,\omega,\vpt)\right] 
	\end{equation}
	is then determined by a rate integral $\Gamma$, with $\partial_t \Gamma = \gamma$.
	
	We compute (\ref{eq:lr_diagonal}) from the quantum evolution equations at NLO large-$N$ using linearization~\cite{Berges:2005ai,Juchem:2003bi} and a lowest-order gradient expansion~\cite{Berges:2005md,Berges:2010ez}. 
	The result for the nonequilibrium response rate is displayed in the right graph of~\rff{fig:decay_wig} for the same time in the universal scaling regime as the spectral function on the left. One observes very similar pronounced structures, taking into account that the response rate is symmetric in frequency with $\gamma(t,\omega,\vpt) = \gamma(t,-\omega,\vpt)$ rather than anti-symmetric as the spectral function. Remarkably, for small momenta the negative regions (red) around $\omega \simeq \pm M$ exhibit the largest values of $|\gamma|$. This has the striking consequence that in the infrared the dominant on-shell rate $\gamma(t,\omega \simeq M,\vpt \simeq 0)$ turns negative.
	According to (\ref{eq:lr_diagonal}) or (\ref{eq:lr_diagonal_solution}), respectively, this gives rise to growing (unstable) perturbations around the universal scaling solution.
	
	The linear response results for the on-shell rate can be compared to the time evolution of perturbations $\delta F(t,t,\vpt)$ obtained from the solution of the full quantum evolution equations at NLO large-$N$. This is shown for various momenta in~\rff{fig:deltaF_comparison}, after an initial ``kick'' $\delta F(t_i,t_i,\vpt)= 10^{-3}\, (t_i/t_{\mathrm{ref}})^{\alpha} F_S\left((t_i/t_{\mathrm{ref}})^\beta \vpt\right)$ at $t_i\,Q=1800$. For the comparison the initial perturbation $\delta F(t_i,t_i,\vpt)$ was fitted as a momentum-dependent constant. One observes that the quasi-exponential growth or decay of the envelopes of oscillations in time is rather well captured by linear response, where the error band indicates a residual mild dependence on the infrared cutoff for the low momenta considered. The rapid oscillation time scale of the full solution is set by the quasi-particle mass~$M(t)$, which is not captured by the linear response analysis employing also a gradient expansion in time. We note that for small enough $\delta F$ also from the full evolution dynamics an approximately linearly independent evolution of the different modes by considering also separately excited modes is observed, which corroborates a linear response approach.
	
	The linear response analysis also allows us to uncover the underlying dynamical scattering processes of the various quasi-particles. This is conveniently described
	by introducing the (off-shell) distribution function $f_p(t)$ with four-momentum $p=(p^0,\mathbf{p})$ according to
	\begin{equation}
	F(t,p) = \left[f_p(t)+\tfrac{1}{2}\right]\rho(t,p)\, .
	\end{equation}
	The leading contribution to the on-shell linear response results arises from elastic (``two-to-two'') scatterings described by
	\begin{eqnarray}
	&&\gamma^{2\leftrightarrow 2}(t,p) = - \frac{\lambda^2}{36 N p^0 }  \int \frac{d^3 q}{(2\pi)^3}\frac{d^3  l}{(2\pi)^3}\frac{d^3 r}{(2\pi)^3}(2\pi)^4 	\label{eq:response_rate_22}\\
	&&\times \int_0^\infty \frac{d q^0 d l^0 d r^0}{(2\pi)^3}\delta^{(4)}(p+l-q-r)
	\rho(t,q)\rho(t,l)\rho(t,r) \nonumber \\
	&&\times \left[v_{\mathrm{eff}}(t,p+l)+v_{\mathrm{eff}}(t,p-q) + v_{\mathrm{eff}}(t,p-r)\right]  \nonumber\\
	&&\times \left[(f_l(t)+1) f_q(t) f_r(t) -f_l(t) (f_q(t)+1) (f_r(t)+1)\right]\nonumber\, .
	\end{eqnarray}
	Here the effective vertex 
	\begin{equation}
	\label{eq:veff}
	v_{\mathrm{eff}}(t,p) = \left|1+\frac{\lambda}{3} \int \frac{d^4 k}{(2\pi)^4} F(t,p-k) G^R(t,k)\right|^{-2}
	\end{equation}
	resums the infinitely many $s$-channel scatterings taken into account at NLO in the large-$N$ expansion depicted in \rff{fig:diagrams}~\cite{Berges:2010ez,Walz:2017ffj}. 
	
	The on-shell response rate $\gamma^{2\leftrightarrow 2}(t,p^0 \simeq M,\vpt \simeq 0)$ in the infrared exhibits two competing contributions of comparable magnitude: a positive (stable) contribution $\gamma^{2\leftrightarrow 2}_+(t,\vp)=\gamma^{2\leftrightarrow 2}(t,p^0=\omega(t,\vp),\vp)$ and a negative (unstable) contribution $\gamma^{2\leftrightarrow 2}_-(t,\vp)=\gamma^{2\leftrightarrow 2}(t,p^0=\omega^-(t,\vp),\vp)$, which are respectively denoted in the right graph of~\rff{fig:decay_wig}. The latter dominates in the scaling regime at $\vpt =0$ for which $\omega(t,\vp=0) = \omega^-(t,\vp=0)=M(t)$. These competing processes can be related to elastic scatterings, where $p^0+\omega(t,\mathbf{l})=\omega(t,\mathbf{q})+\omega(t,\mathbf{r})$, with contributions at $p^0=\omega(t,\vp)$ and additionally at $p^0=\omega^-(t,\vp)$. The latter can be associated to the production of particles with mass $M$ at rest and thereby contribute to the build-up of the macroscopic zero mode.
	
	We observe that the emergence of the competing contributions in this strongly-correlated system hinges on the combination of a gapped dispersion relation and a steep power-law behavior of the distribution function for characteristic infrared momenta $f(t,\vp)\sim \vpt^{-\kappa}$ for $\kappa\gtrsim 2$. In our case $\kappa \simeq 5$ and, in particular, the phenomenon is absent in thermal equilibrium~\cite{Berges:2005ai,Juchem:2003bi,Denicol:2022bsq}. 
	
	\mysection{Universal scaling of perturbations}
	\begin{figure}[t]
		\centering
		\includegraphics[width=1\columnwidth]{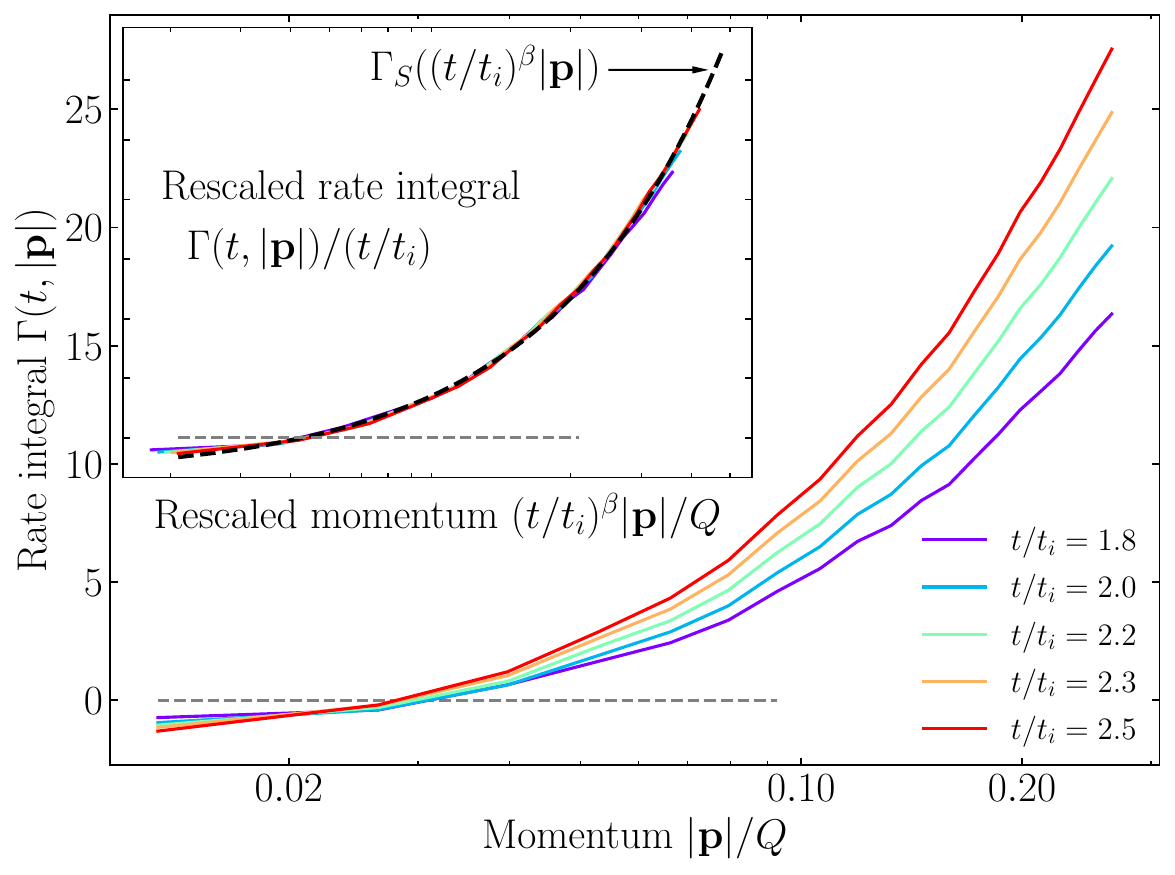}
		\caption{Rate integral as a function of momentum for different times. The inset shows the same data but rescaled, such that all curves at different times collapse to a single scaling function, which is well-described by (\ref{eq:rate_integral_scaling_fct}) as displayed by the dashed black line in the inset.}
		\label{fig:lr_scaling_Gamma}
	\end{figure}
	Since the linear response analysis describes the dynamics of perturbations in terms of the universal scaling solution itself, the response rate $\gamma$ and the respective rate integral $\Gamma$ can exhibit scaling. For the on-shell rate integral we find
	\begin{equation}
	\label{eq:rate_integral_scaling}
	\Gamma(t,\vpt) = (t/t_{\mathrm{ref}})\Gamma_{S}((t/t_{\mathrm{ref}})^{\beta} \vpt)
	\end{equation}
	with scaling function
	\begin{equation}
	\label{eq:rate_integral_scaling_fct}
	\Gamma_S((t/t_{\mathrm{ref}})^\beta \vpt) = A\; (t/t_{\mathrm{ref}})^\beta \vpt/Q - B \, ,
	\end{equation}
	corresponding to an on-shell response rate given by $\gamma(t,\vpt) = (\beta +1) (A/t_{\mathrm{ref}}) (t/t_{\mathrm{ref}})^\beta \vpt/Q - B/t_{\mathrm{ref}}$. 
	
	One observes that the overall scaling exponent of time in~(\ref{eq:rate_integral_scaling}) is unity, which follows analytically from a scaling analysis of~(\ref{eq:response_rate_22}) and~(\ref{eq:veff}). Similarly, it follows that scaling functions do not depend on time and momentum separately but only on the product $t^\beta \vpt$ with universal exponent $\beta$. We confirm these results by analyzing also the numerical solution of the full evolution equations at NLO large-$N$ in~\rff{fig:lr_scaling_Gamma}~\footnote{The grid size was reduced to $N_x=316$ for this figure in order to reach later times.}. While the main graph shows the unrescaled data, the inset demonstrates that the rescaled results all collapse to a single scaling function to very good accuracy.
	
	The form of the scaling function~(\ref{eq:rate_integral_scaling_fct}) is universal up to an overall constant factor and normalization of the argument, which we subsume into the constants $A$ and $B$. Their magnitudes, as well as the universal linear dependence of $\Gamma_S$ on $t^\beta \vpt$, are obtained numerically via a fit of (\ref{eq:rate_integral_scaling_fct}) to the rescaled rate integral, as displayed by the dashed black line in the inset of \rff{fig:lr_scaling_Gamma}. We choose $t_i = t_{\mathrm{ref}}$ and for $t\gg t_i$ we find that $A/(Q t_{\mathrm{ref}})^{\beta+1}=(1.10\pm 0.01) \times 10^{-3}$ and $B/(Qt_{\mathrm{ref}})=(1.41\pm 0.05) \times 10^{-3}$ become independent of the choice of $t_{\mathrm{ref}}$ in units of $Q$, with fit errors indicated. While the values of $A$ and $B$ depend on the chosen microscopic parameters such as coupling and initial conditions, the amplitude ratio $A/B^{\beta+1} = 20.8\pm 1.2$ is universal.
	
	From~(\ref{eq:rate_integral_scaling_fct}) we find that for small enough momenta the negative contributions $\sim B\, t$ to the rate integral $\Gamma$ always dominate. However, for given momentum $\vpt$ and 
	as time progresses the positive (stable) term $\sim A\, t^{\beta+1} \vpt$ will eventually outgrow the negative (unstable) contribution for all non-zero momenta in the scaling regime.

	\mysection{Conclusion}
	Our results establish the stability properties of a nonthermal fixed point from first principles by solving for the dynamics of perturbations around it.
	The perturbations turn out to be captured well by a time- and momentum-dependent response rate $\gamma$. Our results demonstrate that $\gamma$ exhibits universal properties and we determine the dynamical exponents, amplitude ratios and scaling functions. We discover the phenomenon of a scaling instability. While the low-momentum response rate is negative leading to unstable perturbations and higher momenta are stable, $\gamma$ is a self-similar scaling function of~$\sim t^\beta \vpt$. As a consequence, the system shows attractor behavior after~$t \sim 1/|\mathbf{p}|^{1/\beta}$ for any non-vanishing momentum $|\mathbf{p}|$. Since for a system with linear size $L$ the smallest resolved momentum is $\sim 1/L$, measurements in finite systems will always detect attractor properties at late enough times. 
	
	Since the considered universal scaling is known to be present in a wide range of quantum and classical-statistical relativistic as well as non-relativistic many-body systems,
	our results provide an important ab initio example of the emergence of attractor properties in dynamical scaling phenomena. Moreover, the example shows that such ``self-organized'' scaling, in which complexity emerges in a robust way that requires no particular fine-tuning, can be realized in the presence of both stable and unstable directions for the dynamics. This opens new perspectives on the underlying mechanisms and scope of models that have a critical point as an attractor. Moreover, present-day experimental platforms with ultra-cold quantum gases in the many-body regime can give direct access to these intriguing far-from-equilibrium phenomena.
	
	\mysection{Acknowledgments} We thank Gregor Fauth for his involvement during the initial stages of this work. 
	The authors acknowledge support by the state of Baden-Württemberg through bwHPC and the German Research Foundation (DFG) through grant no 
	INST 40/575-1 FUGG (JUSTUS 2 cluster).
	This work is supported by the DFG  under the Collaborative Research Center SFB 1225 ISOQUANT (Project-ID 27381115) and the Heidelberg STRUCTURES Excellence Cluster under Germany's Excellence Strategy EXC2181/1-390900948.
	
	\bibliography{master.bib}
	
\end{document}